\documentstyle[twocolumn,eqsecnum,aps,epsfig]{revtex}

\def\sqr#1#2{{\vcenter{\hrule height.#2pt\hbox{\vrule width.#2pt
height#1pt \kern#1pt \vrule width.#2pt}\hrule height.#2pt}}}

\def\br{\begin{eqnarray}}
\def\er{\end{eqnarray}}
\def\brn{\begin{eqnarray*}}
\def\ern{\end{eqnarray*}}
\def\er{\end{eqnarray}}

\def\r{\rho}

\begin{document}
\draft

\twocolumn[\hsize\textwidth\columnwidth\hsize\csname
@twocolumnfalse\endcsname

\title{{\bf Test compatible metrics and 2-branes}}
\author{Yakov Itin}
\address{Institute of Mathematics,
 Hebrew University of Jerusalem,\\
 Givat Ram, Jerusalem 91904, Israel \ 
  email: itin@math.huji.ac.il}
\maketitle
\widetext
\begin{abstract}
We propose a sufficient condition for a general spherical symmetric static metric to be compatible with classical tests of gravity.
A 1-parametric class of such metrics are constructed. The Schwarzschild metric as well as the Yilmaz-Rosen metric are in this class. By computing the scalar curvature we show that the non-Schwarzschild metrics can be interpreted as close 2-branes. All the manifolds endowed the described metrics   contain in a class of pseudo-Riemannian manifolds with a scalar curvature of a fixed sign.
\end{abstract}

\narrowtext
\vskip2pc]

The resent generalization of AdS/CFT correspondence due to Witten and Yau \cite {W-Y} is based on manifolds with a positive scalar curvature. This result renew an interest to alternative theories of classical gravity. Indeed, all solutions of the vacuum Einstein field equations yield the metric with zero scalar curvature.\\
Einstein's gravitational theory is in a very good accordance with all known experimental data and this fact makes a natural obstacle for modification the foundation of GR. Indeed, almost all alternative theories of gravity are constructed in such a way that they preserve the classical  Schwarzschild solution for a static spherical-symmetric massive particle
\begin{equation}\label{1}
ds^2=(1-\frac m{\r})dt^2-\frac 1{1-\frac m{\r}}d{\r}^2-{\r}^2d\Omega^2.
\end{equation}
On the other hand the most reliable corroborations of GR (such as gravitational red-shift, perihelion procession, bending of light and time delay of radar signals) are based only on two theoretical facts:\\
1) Geodesic Postulate, \\
2) Long distance behavior of  Schwarzschild metric.\\
It is known a rather different metric 
\begin{equation}\label{2}
ds^2=e^{-2\frac mr}dt^2-e^{2\frac mr}(dx^2+dy^2+dz^2), 
\end{equation}
with $r=(x^2+y^2+z^2)^{1/2}$, which is also compatible with the classical macroscopic observation data.\\
The metric (\ref{2}) appeared for the first time in the framework of Yilmaz's scalar theory of gravity  and hardly later in the bi-metric theory of Rosen (see \cite {yilmaz76}, \cite{rose74} and the reference therein). 
We will refer  to (\ref{2}) as Yilmaz-Rosen metric (see, for instance, \cite{Hehl}). Recently the metric  (\ref{2}) was obtained in the Kaniel-Itin model based on a wave-type field equation for a coframe field \cite{K-I}, \cite{it2} and in the Watt and Misner  scalar model of gravity \cite{WM}.\\
Thus there are two different metrics that are in a good accordance with the macroscopic ($r>>m$) observation data. On a small distance $r\simeq m$ these metrics are rather different: the Schwarzschild metric describes the black hole while the Yilmaz-Rosen metric are regular in all points on the manifold (except of the origin).\\
The question in turn is: {\it What is similar in a long-distance analytical behavior of these two metrics and which other metrics have the same behavior?}\\
We work backwards. We do not specify the field equations, but simply postulate a jet-space of solutions which looks promising.\\ 
Consider the general spherical symmetric static metric in spatial conformal (isotropic) coordinates
\begin{equation}\label{3}
ds^2=e^{f(r)}dt^2-e^{g(r)}(dx^2+dy^2+dz^2).
\end{equation}
We are interested in asymptotically flat metrics, thus  require the  functions $f$ and $g$ to have Taylor expansions  of the form 
\br\label{4}
f(r)&=&\frac {a_1}r+\frac{a_2}{r^2}+O(\frac 1{r^3}),\nonumber \\ 
g(r)&=&\frac {b_1}r+\frac{b_2}{r^2}+O(\frac 1{r^3}).
\er
(the zeroth order terms can be vanishing by rescaling the coordinates).
Write the Schwarzschild  line element, in the same isotropic coordinates
\begin{equation}\label{5}
ds^2=\Big(\frac{1-\frac{m}{2r}} {1+\frac{m}{2r}}
\Big)^2dt^2-\Big({1+\frac{m}{2r}}\Big)^4(dx^2+dy^2+dz^2).
\end{equation}
The Taylor expansions of the functions $f$ and $g$ for this metric are
\br\label{6}
f(r)&=&2\ln\frac{1-\frac{m}{2r}} {1+\frac{m}{2r}}=-\frac {2m}r+O(\frac 1{r^3}), \nonumber\\ 
g(r)&=&4\ln({1+\frac{m}{2r}})=\frac {2m}r-\frac{m^2}{2r^2}+O(\frac 1{r^3})
\er
and the coefficients are 
\begin{equation}\label{7}
a_1=-2m, \quad b_1=2m, \quad a_2=0, \quad b_2=-\frac {m^2}2, \ ...
\end{equation}
The corresponding coefficients for the Yilmaz-Rosen metric (\ref{2}) are
\begin{equation}\label{8}
a_1=-2m,\quad b_1=2m,\quad a_i=b_i=0\quad \text{for} \ i>1
\end{equation}
Comparing (\ref{7}) and (\ref{8}) we obtain \it{ a sufficient condition} \rm for a general static spherical-symmetric asymptotic flat metric in isotropic coordinates (\ref{3}) to be compatible with the macroscopic classical tests:
\begin{equation}\label{9}
a_1+b_1=0, \qquad a_2=0.
\end{equation}
In order to describe by the metric (\ref{3}) the field of a point with arbitrary mass, the actual value of the coefficient $a_1$ should, also, be arbitrary. Thus the conditions (\ref{9}) define a 2-jet space of functions $f(r)$ and a 1-jet space of functions $g(r)$. These spaces correspond to the $\frac 32$-order approximation of GR. \\
As an example of functions containing in these jet-spaces consider a family of metrics
\begin{equation}\label{10}
ds^2=\Big(\frac{1-\frac{m}{kr}} {1+\frac{m}{kr}}
\Big)^kdt^2-\Big({1+\frac{m}{kr}}\Big)^{2k}(dx^2+dy^2+dz^2),
\end{equation}
where $k$ is a dimensionless parameter. It is easy to see that the metric (\ref{10}) satisfies the conditions (\ref{10}) for an arbitrary choice of the parameter $k$. \\
In order to have analytically correct functions in (\ref{10}) we require the parameter $k$ to be integer. Note that this restriction is taken only for simplification. We can also consider the parameter $k$ as an arbitrary real number, but in this case we have made an analytical redefinition of the metric on the small distances $r \le \frac mk$.\\
For $k=2$ one obtain, certainly, Schwarzschild metric while in the limits $k\to \pm\infty$   (\ref{10})  approaches Yilmaz-Rosen metric.\\ 
In order to obtain the the metric (\ref{10}) in Schwarzschild coordinates  we have to use  the new radial coordinate $r=r(\r)$, which  is implicit defined by the equation
\begin{equation}\label{11}
\Big({1+\frac{m}{kr}}\Big)^{k}r=\r.
\end{equation}
For $k<0$  the metric (\ref{10}) is singular at a distance
\begin{equation}\label{12} 
r=-\frac mk \quad ==>\quad \r=0
\end{equation}
i.e. in the origin of the Schwarzschild coordinates.\\
As for $k>0$ the metric (\ref{8}) is singular on a sphere
\begin{equation}\label{13} 
r=\frac mk \quad ==>\quad \r=m\frac {2^k}k.
\end{equation}
Note that the physical radius of singular sphere $\r$ increases very fast with growth of the parameter $k$. 
In order to clarify the nature of  the singularities compute 
the scalar curvature of the metric (\ref{10})  
\begin{equation}\label{14}
R=\frac{2-k}{k}\bigg(\frac{m^2}{r^4}\cdot\frac{1+(1-\frac{m}{kr})^2}{(1-\frac{m^2}{k^2r^2})^2}\Big(1+\frac{m}{kr}\Big)^{-2k}\bigg).
\end{equation}
Note that the expression in the brackets are positive thus the sign of the scalar curvature depends only on the value of the parameter $k$. Thus all the manifolds endowed the metric (\ref{10})  contain in a class of pseudo-Riemannian manifolds with a scalar curvature of a fixed sign.\\
The scalar curvature is zero only in the case of Schwarzschild metric  -  $k=2$. \\
Consider the different regions for the values of the parameter $k$:\\
${\mathbf 1)\qquad k> 2}$\\
The scalar curvature negative in every final point on the manifold.
Near the singular value of coordinates $r=\frac mk$ the scalar curvature is singular $R\to -\infty$. Thus this coordinate singularity is physical. The scalar curvature inside of the spherical envelope decreases very fast. 
This singularity can be interpreted as a rigid sphere - close 2-brane.

\begin{figure}[t]
\label{fig:1}
\begin{center}
\epsfxsize=2.25in
\includegraphics[angle=270,width=8cm]{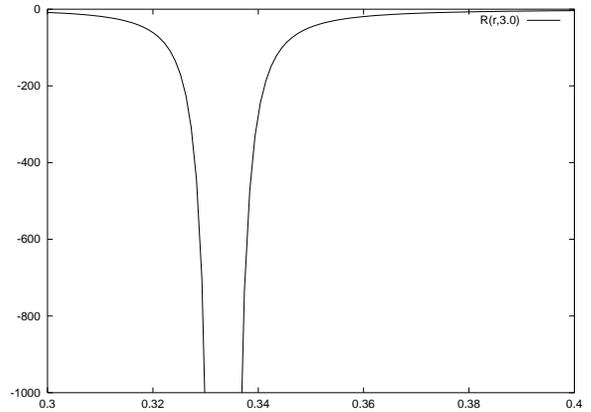}
\end{center}
\caption{
The scalar curvature (\ref{14}) plotted as a function $R/m^2$ of a radial distance $r/m$ for  $k=3$.
}
\end{figure}
For  $k\to \infty$ we obtain the expression for the scalar curvature of Yilmaz-Rosen metric \cite{K-I}
\begin{equation}\label{15}
 R= -2\frac{m^2}{r^4}e^{-2\frac mr}.
\end{equation}
\\
${\mathbf 2)\qquad k=1}$\\
The scalar curvature is positive.
We have the metric in isotropic coordinates 
\begin{equation}\label{16} 
ds^2=\Big(\frac{1-\frac{m}{r}} {1+\frac{m}{r}}
\Big)dt^2-\Big({1+\frac{m}{r}}\Big)^{2}(dx^2+dy^2+dz^2).
\end{equation}
From the relation (\ref{11}) we obtain the transform to the Schwarzschild radial coordinate
$$\r=r+m.$$
The metric in these coordinates takes the form
\begin{equation}\label{17} 
ds^2=(1-2\frac m{\r})dt^2-\frac 1{(1-\frac m{\r})^2}d{\r}^2-{\r}^2d\Omega^2.
\end{equation}
The scalar curvature of the metric (\ref{16}) is 
\begin{equation}\label{18}
R=\frac{m^2}{r^4}\cdot\frac{1+(1-\frac{m}{r})^2}{(1-\frac{m}{r})^2(1+\frac{m}{r})^{4}}.
\end{equation}
This expression is positive in every point of the manifold. It is singular for $r=m$ or equivalently for $\r=2m$. Thus the coordinate singularity for $\r=2m$ is physical. 
\begin{figure}[t]
\label{fig:2}
\begin{center}
\epsfxsize=2.25in
\includegraphics[angle=270,width=8cm]{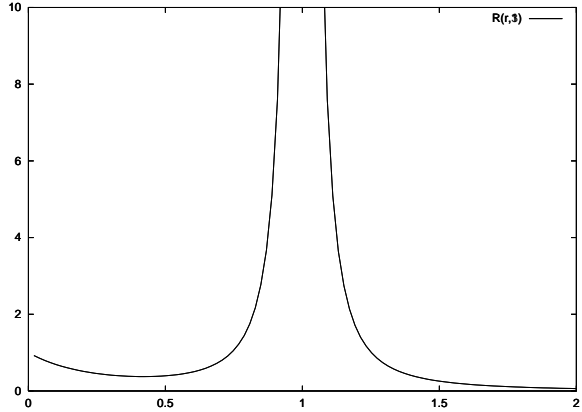}
\end{center}
\caption{
The scalar curvature (\ref{14}) plotted as a function $R/m^2$ of $r/m$ for  $k=1$.
}
\end{figure}
This singularity can be also interpreted as a rigid sphere - close 2-brane.\\
${\mathbf 3)\qquad k\le -1}$\\
The scalar curvature is negative.
Consider for instance $k=-1$. 
The metric takes the form
\begin{equation}\label{19}
ds^2=\Big(\frac{1-\frac{m}{r}} {1+\frac{m}{r}}
\Big)dt^2-\Big({1-\frac{m}{r}}\Big)^{-2}(dx^2+dy^2+dz^2).
\end{equation}
The metric is singular at the  Schwarzschild radius $r=m$
The scalar curvature takes a form
\begin{equation}\label{20}
R=-{3}\frac{m^2}{r^4}\cdot\frac{1+(1+\frac{m}{r})^2}{(1+\frac{m}{r})^2}.
\end{equation}
This value is regular for every $r$ (except of the origin). 
\begin{figure}[t]
\label{fig:3}
\begin{center}
\epsfxsize=2.25in
\includegraphics[angle=270,width=8cm]{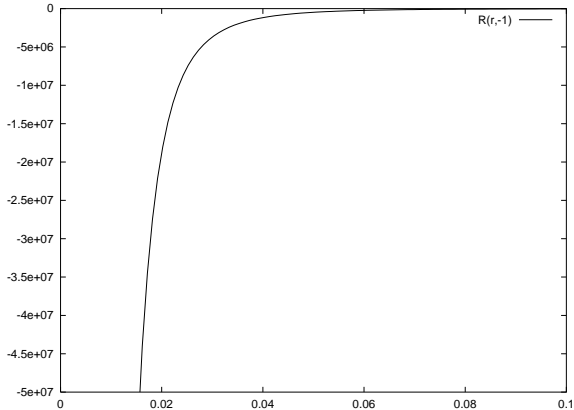}
\end{center}
\caption{
The scalar curvature (\ref{20}) plotted as a function $R/m^2$ of $r/m$ for  $k=1$.
}
\end{figure}
In order to describe by the metric (\ref{19}) a black hole one should locate it's surface at a distance $r=m$ where $g_{00}=0$. In fact this surface cannot be reached by any material object. The proper radial distance from the surface $r=m$ to a point $r_0>m$ is 
$$l=\int^{r_0}_m{\frac {dr}{(1-\frac mr)^2}}\to \infty$$
The proper time for a radial null geodesic is also infinite
$$T=\int^{r_0}_m \frac {(1+\frac mr)^\frac 12}  {(1-\frac mr)^\frac 32}dr\to \infty.$$
Thus the surface of a star cannot never reach the   Schwarzschild radius and a realistic physical system can be modeled by the metric (\ref{19}) only for a distance $r>m$. The behavior of the metric is similar to the spherical-symmetric solution in the gravity model of Lee and Lightman. \cite{L-L}.

\end{document}